\documentclass[]{pasj01}
\begin{document}  

\def\Msun{M_{\odot \hskip-5.2pt \bullet}}   
\def\r{\bibitem[]{}}  \def\kms{km s$^{-1}$}  \def\deg{^\circ}   
\def\Htwo{H$_2$\ }  \def\fmol{f_{\rm mol}} \def\Fmol{ $f_{\rm mol}$ }
\def\rhoh2{\rho_{\rm H_2}} \def\rhohi{\rho_{\rm HI}} \def\rhotot{\rho_{\rm tot}}\def\rhot{\rho_{\rm tot}}
\def\alphah2{\alpha_{\rm H_2}} \def\alphat{\alpha_{\rm tot}}
\def\Sigmat{\Sigma_{\rm tot}}\def\Sigmatot{\Sigma_{\rm tot}} 
\def\Sigmah2{\Sigma_{\rm H_2}} \def\Sigmahi{\Sigma_{\rm HI}} 
\def\sigmahii{\Sigma_{\rm HII}} \def\CC{${\rm cm}^{-3}$}
\def\sfu{\Msun~{\rm y^{-1}~kpc^{-2}}} \def\sfuvol{\Msun~{\rm y^{-1}~kpc^{-3}}}
\def\log{{\rm log}}
\def\hcc{{\rm H~cm^{-3}}} \def\Hcc{ $\hcc$ }\def\Htot{ H$_{\rm tot}$ } 
\def\ssfr{\Sigma_{\rm SFR}} \def\vsfr{\rho_{\rm SFR}} 

\title{Star Formation Law in the Milky Way}  
\author{Yoshiaki \textsc{Sofue}\altaffilmark{1} and Hiroyuki \textsc{Nakanishi}\altaffilmark{2} }
\altaffiltext{1}{Insitute of Astronomy, The University of Tokyo, Mitaka, Tokyo 181-0015, Japan } 
\altaffiltext{2}{Graduate Schools of Science and Engineering, Kagoshima university, 1-21-35 Korimoto, Kagoshima 890-8544, Japan}
\email{sofue@ioa.s.u-tokyo.ac.jp}
@
\KeyWords{galaxies: individual (Milky Way) --- ISM: HI gas --- ISM: molecular gas --- ISM: HII regions --- ISM: star formation }

\maketitle  

\begin{abstract} 
The Schmidt law (SF law) in the Milky Way was investigated using 3D distribution maps of HII regions, HI and molecular (\Htwo) gases with spatial resolutions of $\sim 1$ kpc in the Galactic plane and a few tens of pc in the vertical direction. HII regions were shown to be distributed in a star-forming (SF) disk with nearly constant vertical full thickness 92 pc in spatial coincidence with the molecular gas disk.
The vertically averaged volume star formation rate (SFR) $\vsfr$ in the SF disk is related to the surface SFR $\ssfr$ by
$\vsfr /[\sfuvol] =9.26\times \ssfr/[\sfu]$.
The SF law fitted by a single power law of gas density in the form of 
$\ssfr \propto \vsfr \propto \rho_{\rm gas}^\alpha$ and 
$\propto \Sigma_{\rm gas}^\beta$ 
showed indices of
$\alpha=0.78 \pm 0.05$ for $\rho_{\rm H_2}$ and 
$2.15 \pm 0.08$ for $\rho_{\rm total}$, and
$\beta=1.14\pm 0.23$ for $\Sigma_{\rm total}$, 
where $\rho$ and $\Sigma$ denote volume and surface densities, respectively.
The star formation rate is shown to be directly related to the molecular gas, but indirectly to HI and total gas densities. The dependence of the SF law on the gaseous phase is explained by the phase transition theory between HI and \Htwo gases.
\end{abstract}    
 
\section{INTRODUCTION}

    The Schmidt (Schmidt-Kennicutt) law (hereafter star formation or SF law) linking the star formation rate (SFR) with surface density of interstellar gas by a power law has been established as the universal scaling relation for many galaxies (Schmidt 1959; Kennicutt 1998a, b; Komugi et al. 2006; Lada et al. 2012; Kennicutt and Evans 2012; and the large number of papers cited therein). 
     In the Milky Way, however, the SF relation has been obtained for individual star forming regions, clouds, and/or restricted regions and arms (Fuchs et al. 2006; Luna et al. 2006; Heiderman et al. 2010; Willis et al. 2015; Heyer et al. 2016; and the literature therein), but the law in the entire Galaxy, comparable to that for galaxies, seems to have been not obtained.
 
 Star formation in the Galactic disk is regulated by sequential transition of interstellar gas from HI to molecules followed by contraction of clouds to form stars. The current SF law for surface density of gas, therefore, includes the effect of phase transition from HI to molecules, which are not related linearly (Elmegreen 1993; Sofue and Nakanishi 2016).  
 
 This paper aims at deriving the global SF law in the entire Galaxy, both for volume and surface densities, and clarifies effects dependent on the gas phases of HI and \Htwo. We use the catalogue of HII regions (Hou and Han 2014) and the 3D maps of HI and \Htwo gases (Nakanishi and Sofue 2005, 2006, 2016) which were constructed from the Leiden-Argentine-Bonn All Sky HI survey (Kalberla et al. 2005) and the Galactic disk CO line survey (Dame et al. 2001). Surface and volume SFR are denoted by $\ssfr$ and $\vsfr$, respectively, and 'SFR' in the text and figures means $\ssfr$, unless commented.
 
\section{Distributions of HII Regions and Interstellar Gas}

\subsection{Star forming disk}

\begin{figure} 
\begin{center}      
\includegraphics[width=8cm]{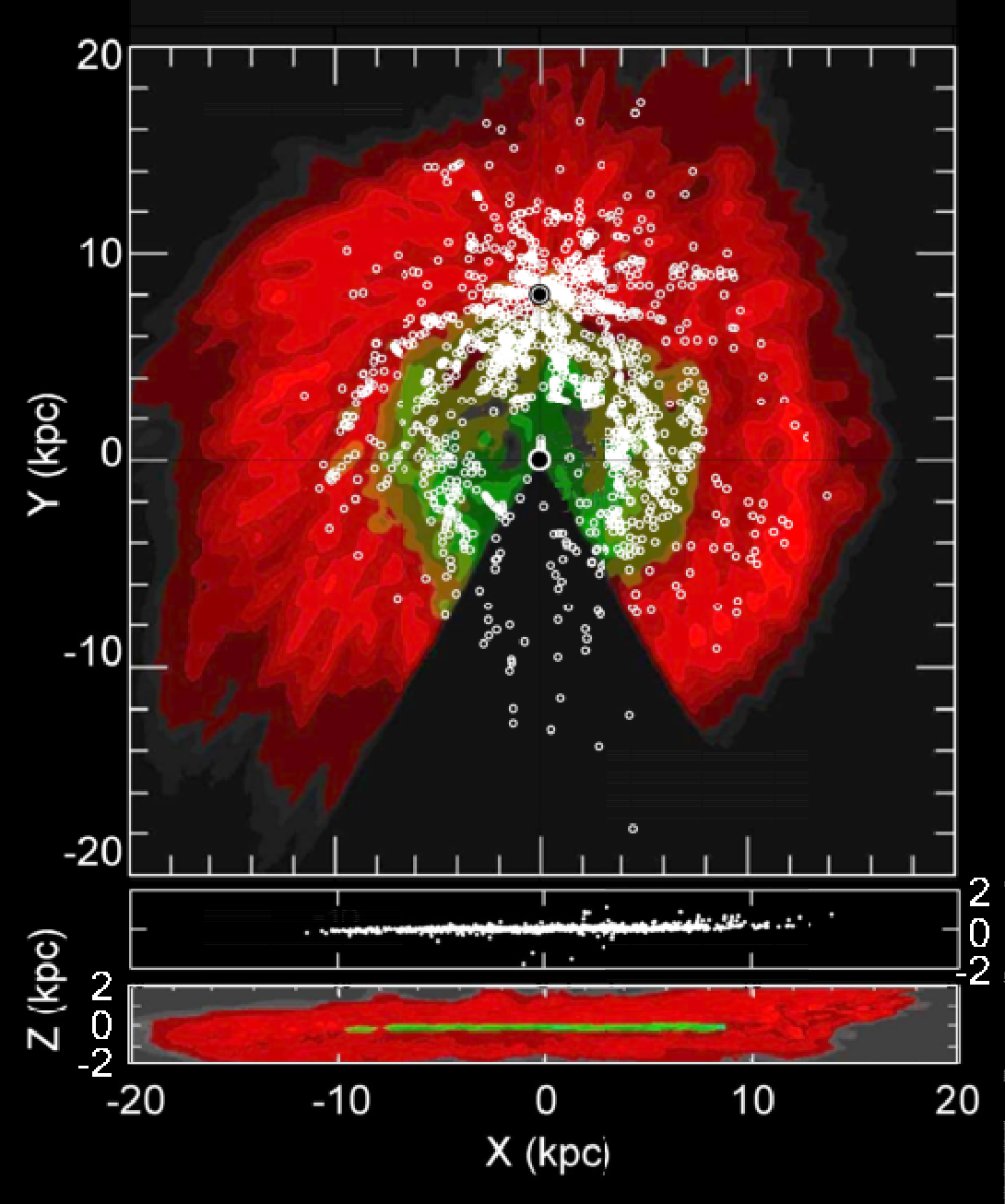}  
\end{center}
\caption{Distribution of column (surface) densities of \Htwo (green) and HI (red) gases in the Galaxy in the ($X,Y$) plane and on the ($X,Z$) plane (Nakanishi and Sofue 2005, 2006, 2016). Positions of HII regions (Hou and Han 2014) are overlaid. As to gas density scale.}
\label{HIIplot}  
\end{figure}

Figure \ref{HIIplot} shows the distribution of HII regions made from the
catalogue by Hou and Han (2014). The distances are scaled
to the galacto-centric distance of the Sun at 8.0 kpc instead of
their 8.5 kpc. The HII regions are superposed on the map of
surface densities of HI (red) and H2 (green) gases projected on
the galactic (X;Y ) plane. Also shown are the projection of the
HII regions on the (X,Z) plane superposed on the cross section
of the gaseous disk (Nakanishi and Sofue 2005-2016).

\begin{figure} 
\begin{center}      
\includegraphics[width=8cm]{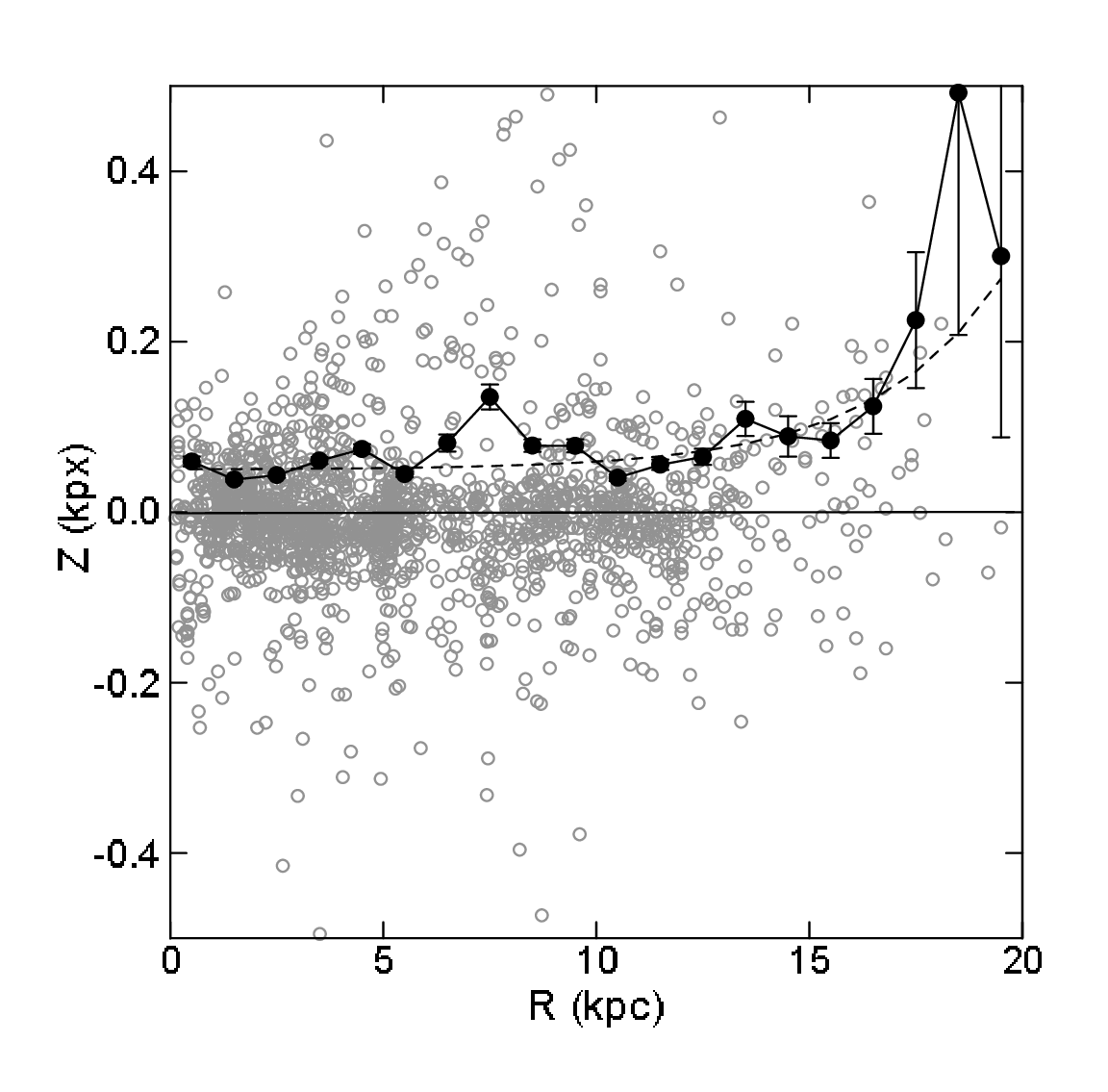}  
\end{center}
\caption{Distribution of HII regions in $R-Z$ plane (grey dots). Black circles are mean absolute values of $Z$ averaged in 1 kpc interval of radius. The dashed line is an approximate fit to the mean height.  }
\label{RZ} 
%\end{figure} 
%\begin{figure} 
\begin{center}      
\includegraphics[width=8cm]{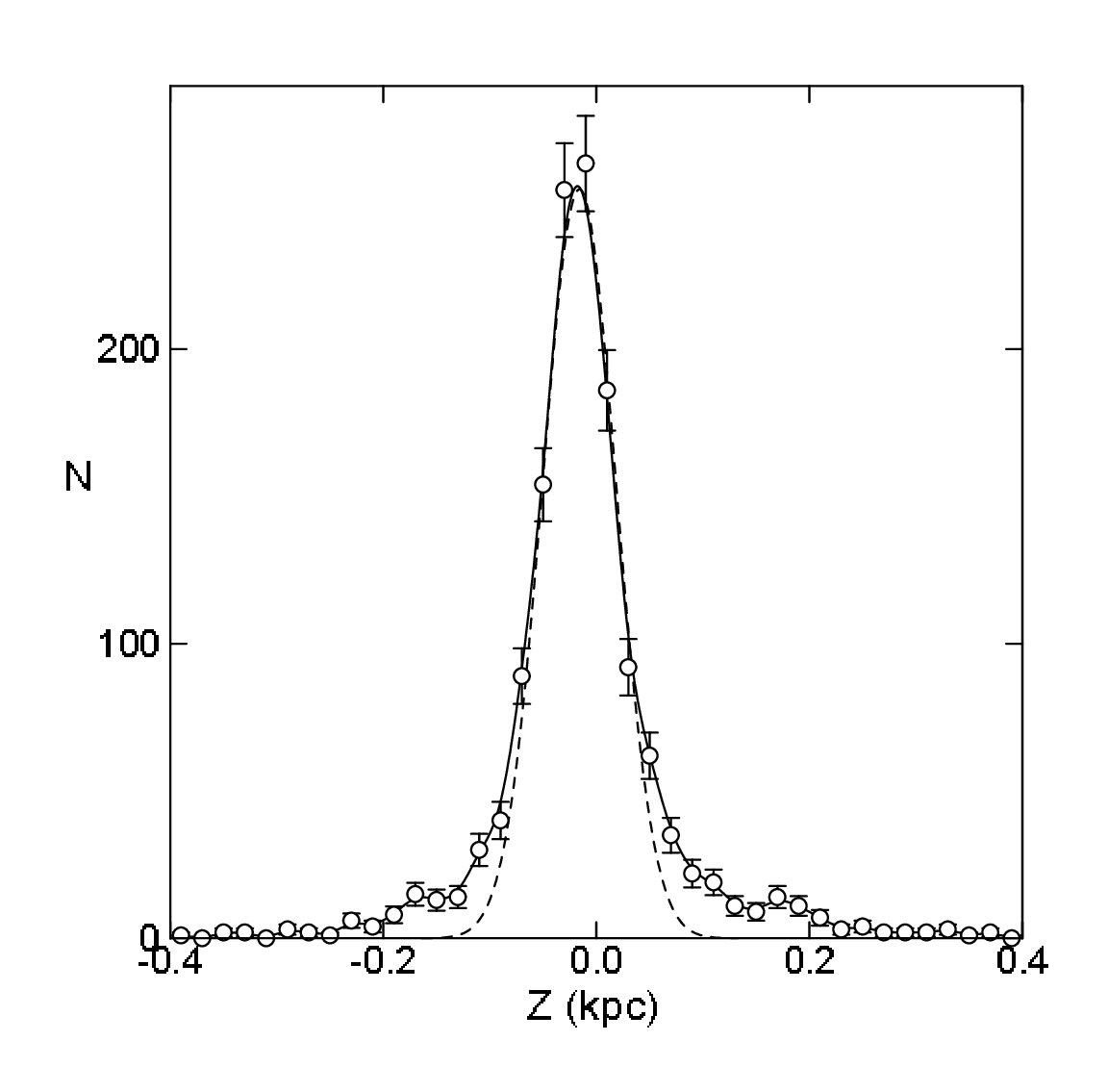}  
\end{center}
\caption{Number of HII regions $N$ per $Z$ interval of 0.02 kpc in $1\le R \le 10$ kpc  (open circles with error bars by $\pm \sqrt{N}$). The dashed curve is an approximate fit to the plot by a Gaussian function of $e$-folding scale thickness $h=0.046$ kpc. }
\label{NZ} 
\end{figure}

Both the HII and gas distributions were obtained for similar rotation curves, which are almost the same inside the solar circle and are similarly flat in the mapped region of the HII regions. Thus, the kinematical distances of HII regions and the gas coincide within errors of a few hundred pc. This accuracy is enough for the present analyses at a typical resolution of $\sim \pm 0.5$ kpc, 

Figure \ref{RZ} shows the distribution of HII region in the $(R, Z)$ plane, and mean values of absolute heights averaged in every 1 kpc interval of radius. The $Z$ distribution is well represented by a flat disk with scale height $h\simeq 0.05$ kpc with outer scale height increasing rapidly as 
\begin{equation}
h\simeq 0.05 (1.+{\rm exp}((R-15{\rm kpc})/3{\rm kpc}))\ [{\rm kpc}].
\end{equation}  

In order to determine $h$ more precisely, we plot in figure \ref{NZ} the vertical distribution of number of HII regions at $1 \le R \le 10$ kpc.
The distribution can be fitted by a Gaussian distribution of $e$-folding scale height of $h=46$ pc, and hence full thickness of $2h=92$ pc. This is in good coincidence with the molecular gas full thickness of 90 to 100 pc inside the solar circle (Nakanishi and Sofue 2006; Heyer and Dame 2015). 

Since the thickness of the disk of HII regions is nearly
constant, the surface SFR per unit area, $\ssfr$,
can be related to the volume SFR, $\vsfr$, by
\begin{equation}
\vsfr=\ssfr/(2h),
\end{equation}
or 
\begin{eqnarray} 
\vsfr[\Msun \ {\rm y^{-1} kpc^{-3}}] 
= 10^9 \vsfr[\Msun \ {\rm y^{-1} pc^{-3}}]\nonumber \\
%\simeq 9.26 \Sigma_{\rm SFR}[\Msun \ {\rm y^{-1} \ kpc^{-2}}]
= 9.26 \ssfr [\Msun \ {\rm y^{-1} \ kpc^{-2}}].
\label{eqSFRvolsurf}
\end{eqnarray} 
 
\subsection{Definition of SFR}

Since the Galactic disk is opaque to $H_\alpha$ emission which is commonly used to approximate SFR, we here employ UV luminosity $L$ of HII regions estimated by using excitation parameter $U$ measured by radio observations (Schraml and Mezger 1969; Hou and Han 2014). We define the surface SFR by
\begin{equation}
\ssfr=M_* \tau_*^{-1} \Sigma_{\rm HII}\left(L\over L_0\right),
\label{eq_SFR}
\end{equation}
where $M_*$ is the mean stellar mass of HII region, $\tau_*$ is the mean life time approximated by the lifetime of involved OB and A stars, $\Sigma_{\rm HII}$ is the surface density of HII regions on the Galactic disk, and $L_0$ is the representative value of $L$. Luminosity $L$ is proportional to emission measure $EM$ times area $s$ of HII region, as $L\sim EM s$ 

Remembering that $EM\sim n_{\rm e}^2 r$ [pc cm$^{-6}$] and $s\sim \pi r^2$ [pc$^2$], $L$ is related to the excitation parameter $U$ [pc cm$^{-2}]$ as
\begin{equation}
{L \over L_0} \simeq \left(U \over U_0 \right)^3,
\end{equation}
where suffix 0 denotes the representative (mean) values. 
Assuming that $M_*=10^4 \Msun$ and $\tau_*=10^7$ y, and $U_0=50$ pc cm$^{-2}$ (Hou and Han 2014), SFR is written as 
\begin{equation}
{\ssfr\over \Msun\ {\rm y^{-1}\ kpc^{-2}}}
=1.0\times 10^{-3}\eta \left(\Sigma_{\rm HII} \over{\rm kpc}^{-2}\right) 
\left(U \over 50 {\rm pc\ cm^{-2}} \right)^3.
\end{equation}

Here, $\eta$ is a calibration factor for a case when more precise values of $M_*$, $\tau_*$ and $U_0$ become available, and is given by
\begin{equation}
\eta={M_* \over 10^4 \Msun} \left(\tau_* \over 10^7{\rm y} \right)^{-1} \left(U_0 \over 50 {\rm pc\ cm^{-2}} \right)^{-3}.
\end{equation}
Since precise determination of $\eta$ factor is beyond the scope of this work, as we are aiming at determining the power-law index of SF law, we assume $\eta=1$ in this paper.

\subsection{Correction of incompleteness}

Nearer HII regions are more easily detected, while detection is
limited in distant sources. Hence, less luminous HII regions
tend to be more crowded near the Sun as seen in figure \ref{HIIplot}. In order to minimize such incompleteness of sampling due to detection limits of observations, we correct the SFR for the detection probability $P$ of HII regions as
\begin{equation}
\ssfr=\ssfr({\rm raw})/P,
\end{equation}
where $\ssfr({\rm raw})$ denotes the observed (detected) raw numbers of HII regions. We assume that $P$ is inversely proportional to flux density, and hence related to distance by
\begin{equation}
P={1 \over 1+(d/d_{\rm c})^2}.
\end{equation}
Here, $d_{\rm c}$ is the distance within which the detection probability for a source with luminosity $L$ is unity, and is given here as
\begin{equation}
d_{\rm c}=d_0[1+(L/L_0)^{1/2}].
\end{equation}
This implies that $P$ for a brighter source than $L_0$ decreases as $\sim L d^{-2}$, while it is almost unity within $d_0$ regardless the luminosity. We here assume that $d_0=3$ kpc.
  
\subsection{Distributions of SFR, H$_2$, HI and total gas densities} 

\begin{figure*} 
\begin{center}   
\includegraphics[width=17cm]{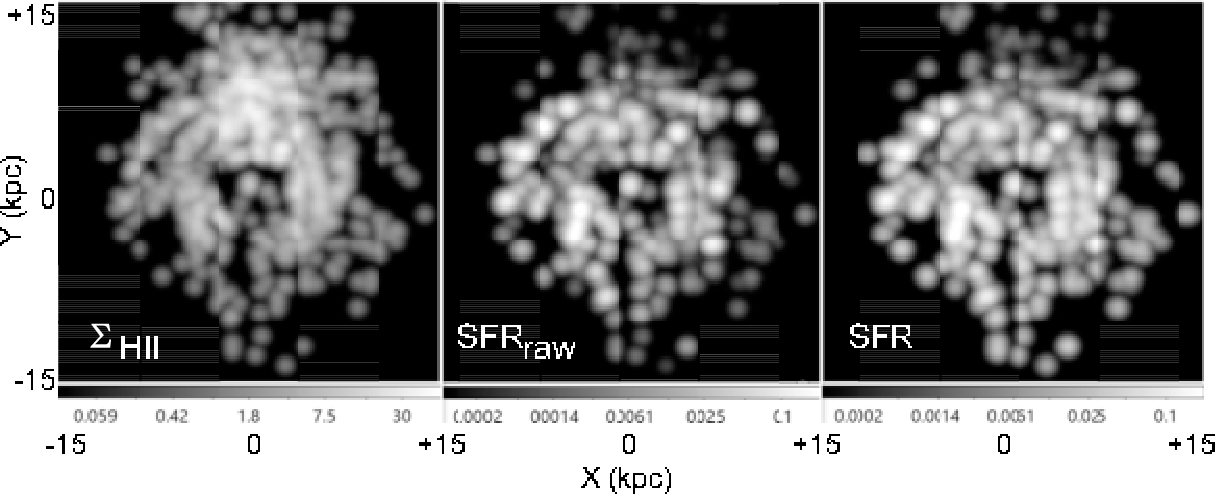}  
\end{center} 
\caption{Left: Distribution of surface number density of HII regions $\Sigma_{\rm HII}$ (same as figure \ref{HIIplot}, but Gaussian smoothed) by logarithmic scaling in unit of number kpc$^{-2}$ as indicated by the bar. Middle: Raw SFR ($\Sigma_{\rm HII}\times$luminosity) by logarithmic intensity scaling in unit of $\Msun {\rm y^{-1} kpc^{-2}}$. Right: SFR, same as middle but corrected for detection probability. } 
\label{XY_SFR} 

\begin{center}
\includegraphics[width=17cm]{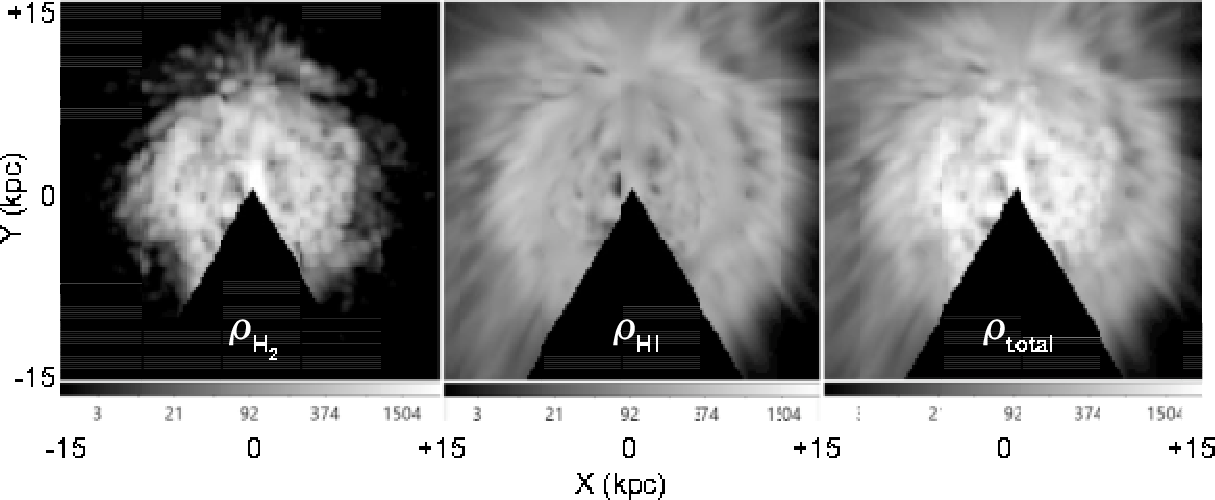}  
\end{center}
\caption{  Distributions of volume densities $\rho$ of H$_2$, HI and total gas density. Intensity scale is logarithmic in unit of $0.01\{\rm H \ cm^{-3}$.}
\label{XY_rho} 

\begin{center}
\includegraphics[width=17cm]{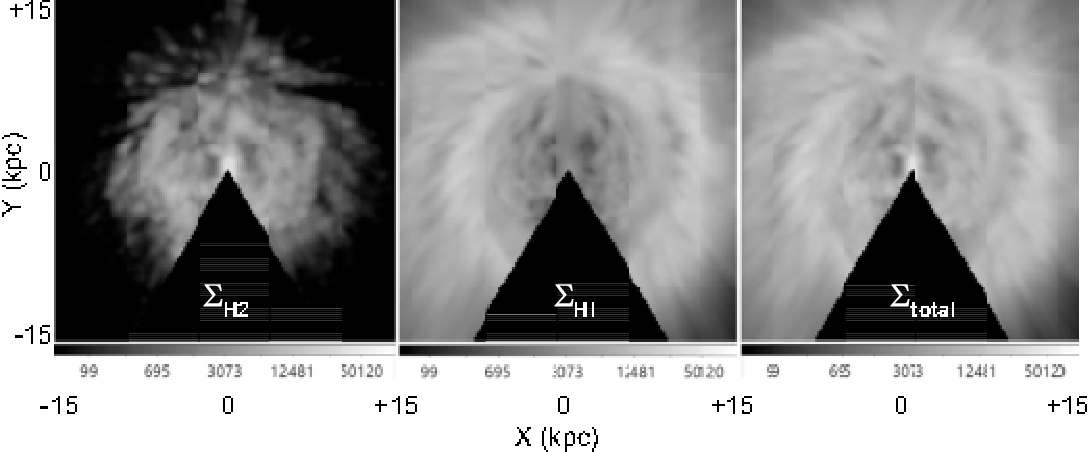}  
\end{center}
\caption{ Distributions of integrated column (surface) densities $\Sigma$ of H$_2$, HI and total gas density. Intensity scale is logarithmic in unit of $1.0\times 10^{-3}\Msun {\rm pc^{-2}}$.}     
\label{XY_Sigma}   
\end{figure*}

We constructed a map of surface distribution of the SFR in the entire Galaxy as shown in figure \ref{XY_SFR}, where are shown the distribution of number density of HII regions (left), the raw SFR as defined by equation \ref{eq_SFR} (middle), and corrected SFR for the detection probability (right). The map was obtained by Gaussian smoothing of SFR calculated for individual HII regions in figure \ref{HIIplot} at $0.2 \times 0.2$ kpc grids with a Gaussian beam of FWHM 1 kpc ($\pm 0.5$ kpc). It is shown that the corrected SFR is more smoothly distributed over the Galactic disk. 

The distribution of SFR may be compared with the distribution maps of neutral gases in the Galaxy as shown in figures \ref{XY_rho} and \ref{XY_Sigma}, which show volume densities in the galactic plane and integrated surface densities, respectively, individually for H$_2$, HI and total gases. The spatial resolution in these maps is 0.5 to 1 kpc in the galactic plane, depending on the distance and direction from the Sun for their kinematical distances. The vertical resolution of the original 3D cubes of HI and \Htwo maps is about 20 to 50 pc, while we here discuss the volume densities near the galactic plane, and the surface densities projected on the galactic plane.

As shown later by Schmidt plots, the clearest correlation is seen between the SFR and the volume and surface densities of the molecular gas. On the other hand, there appears little correlation between the SFR and HI densities.\\

\section{Star Formation Law} 

\begin{figure*}   
\begin{center} 
$\ssfr-\rho$\\
\hskip -5mm\includegraphics[width=5.9cm]{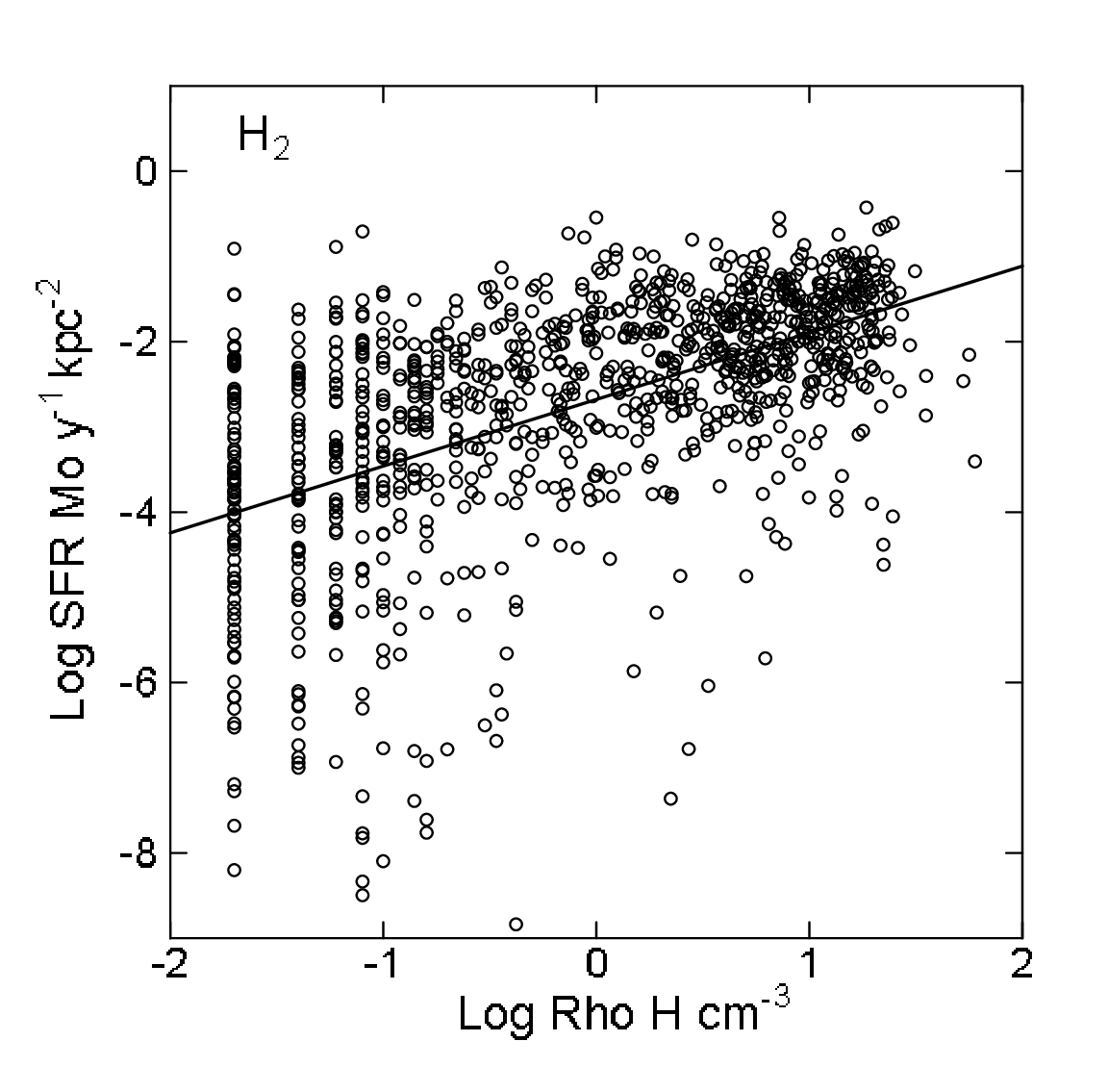}
\hskip -4mm\includegraphics[width=5.9cm]{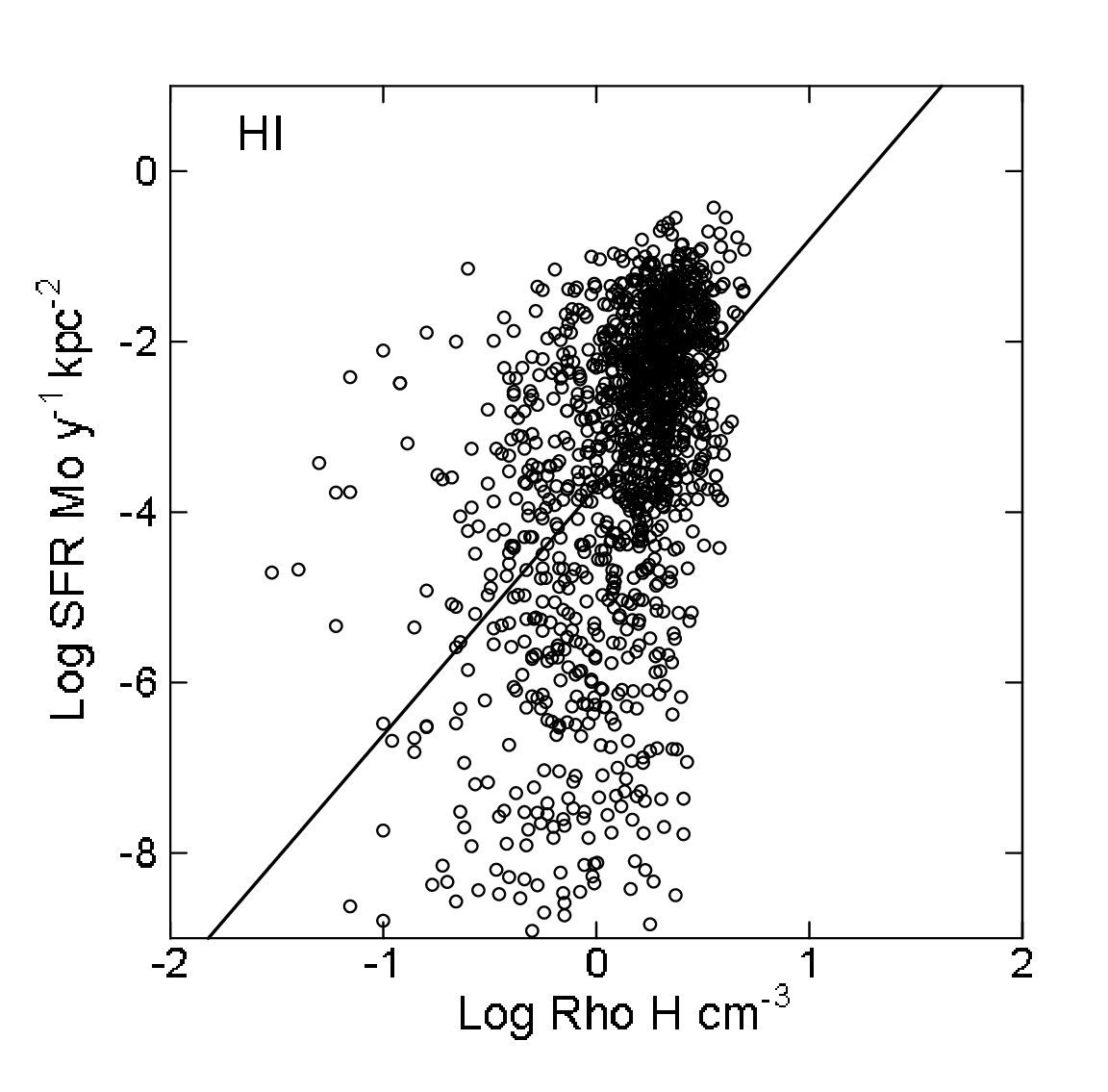}
\hskip -4mm\includegraphics[width=5.9cm]{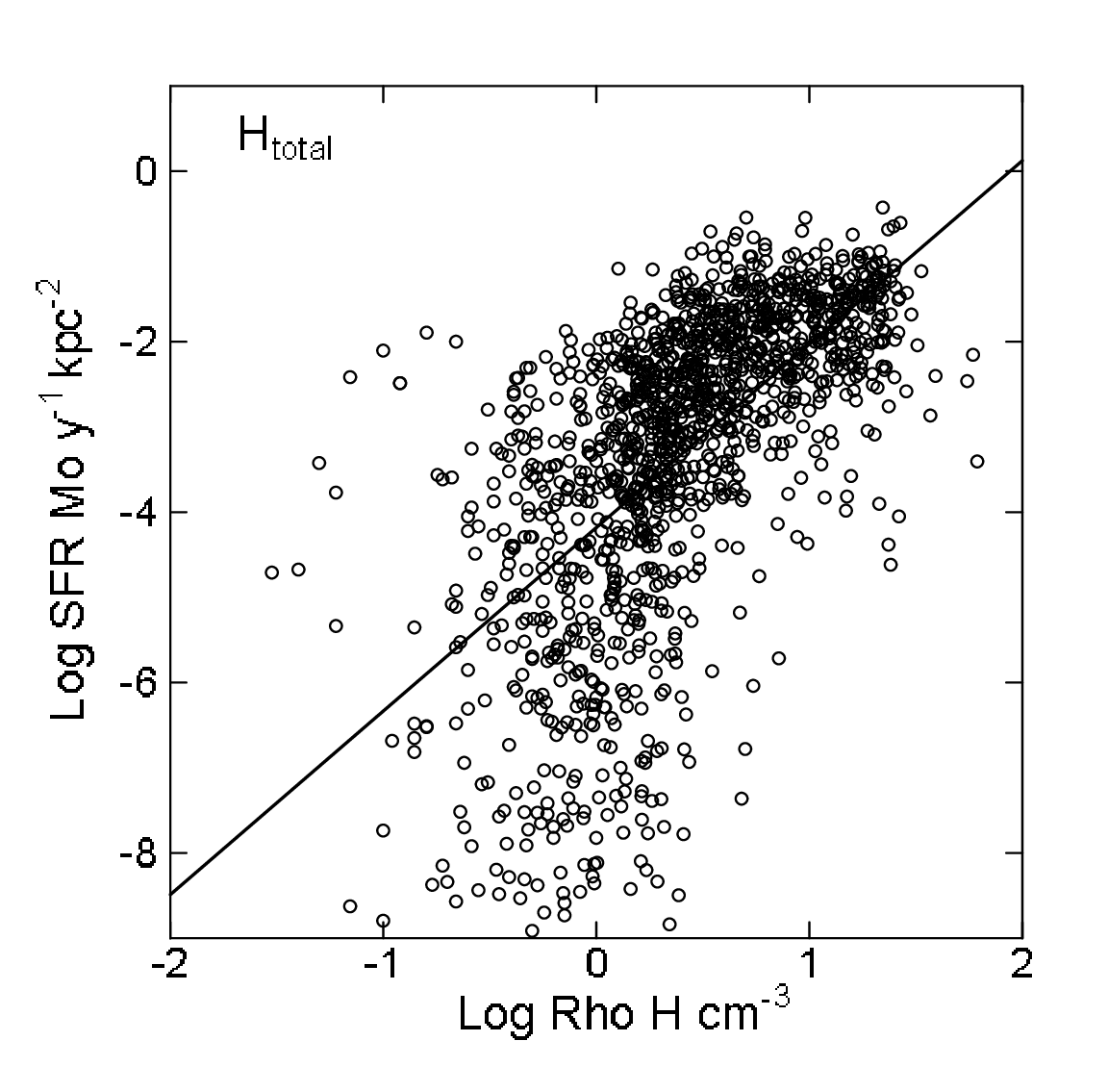}\\
$\ssfr-\Sigma$\\
\hskip -5mm\includegraphics[width=5.9cm]{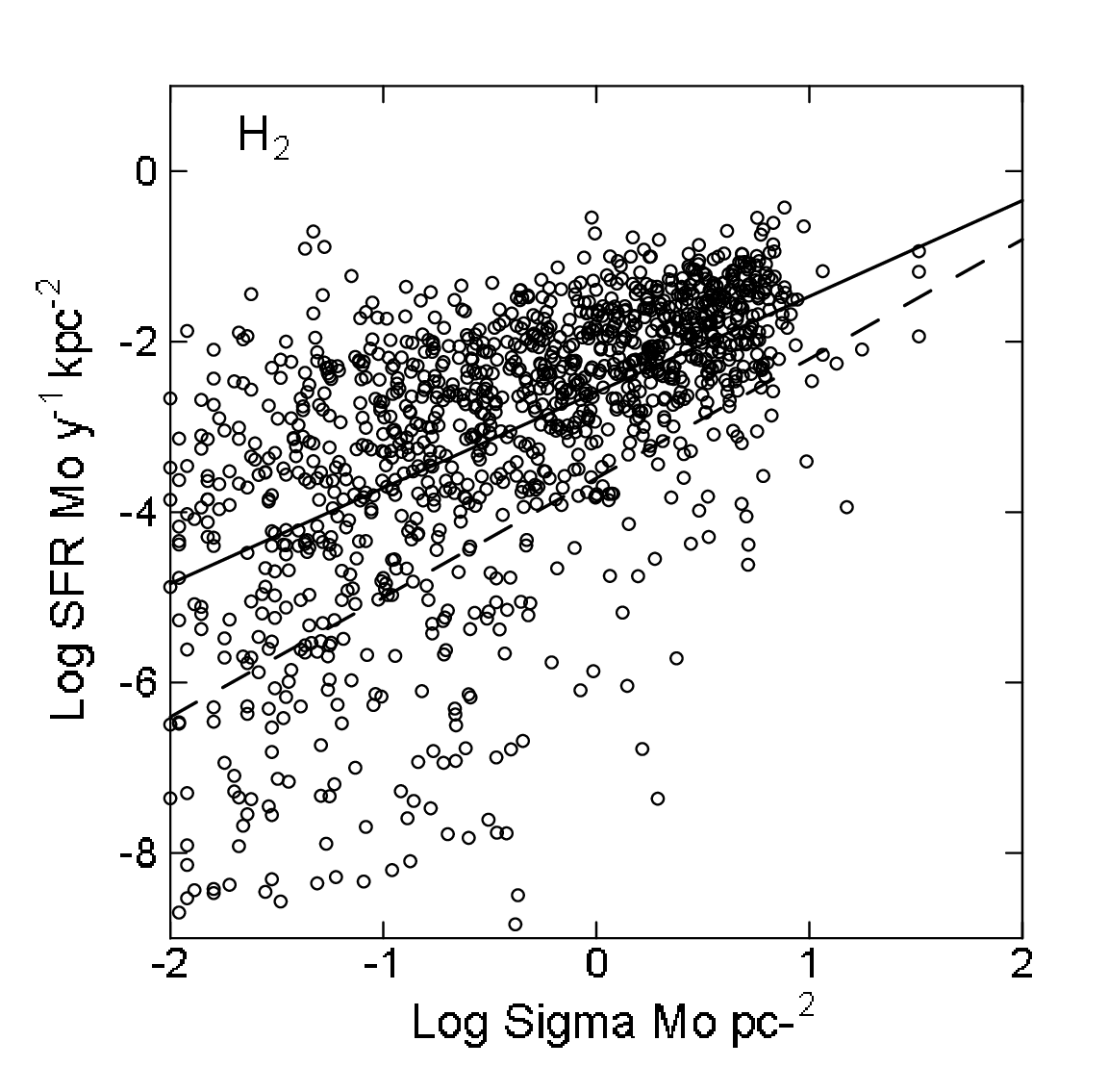}
\hskip -4mm\includegraphics[width=5.9cm]{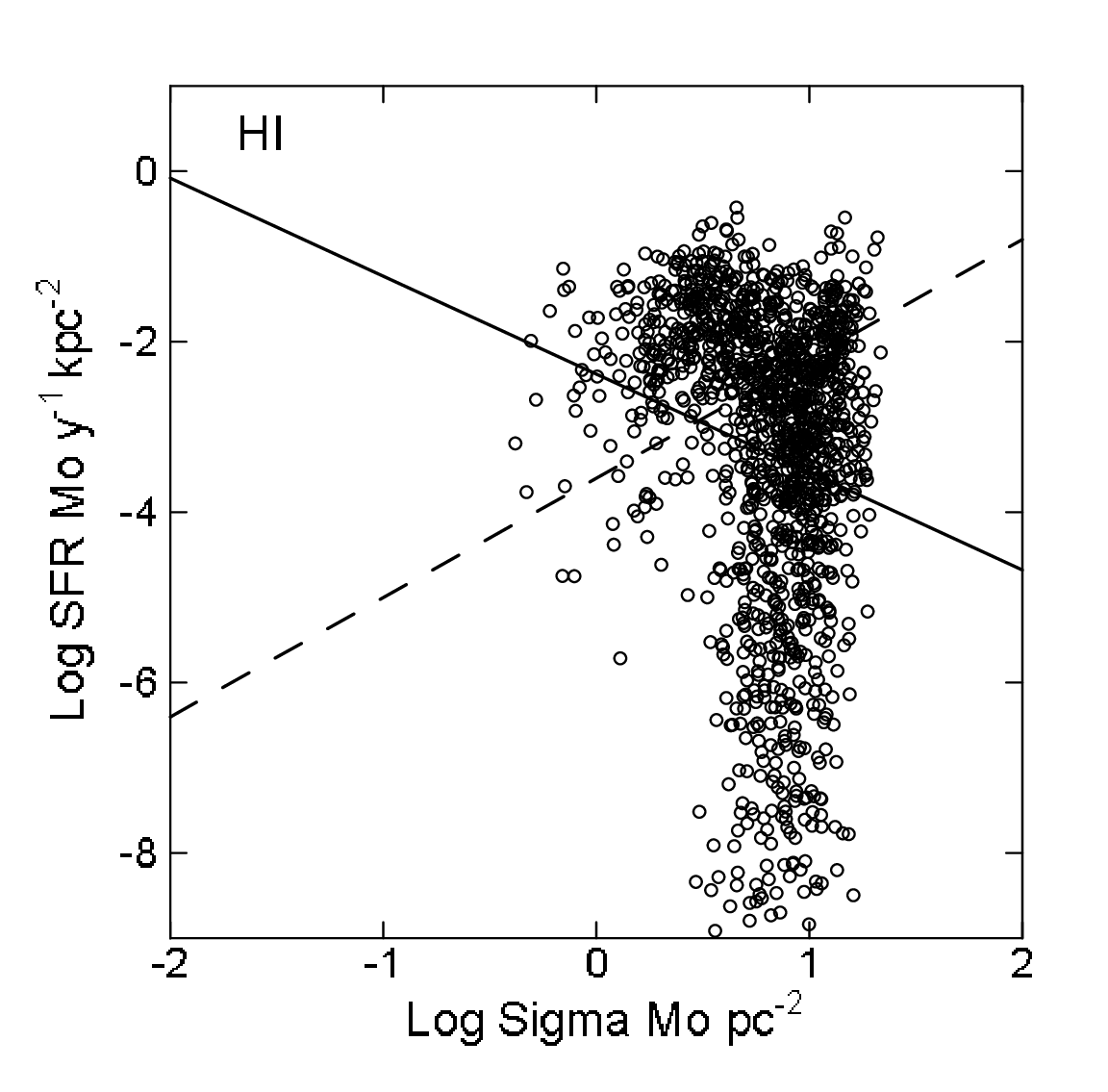}
\hskip -4mm\includegraphics[width=5.9cm]{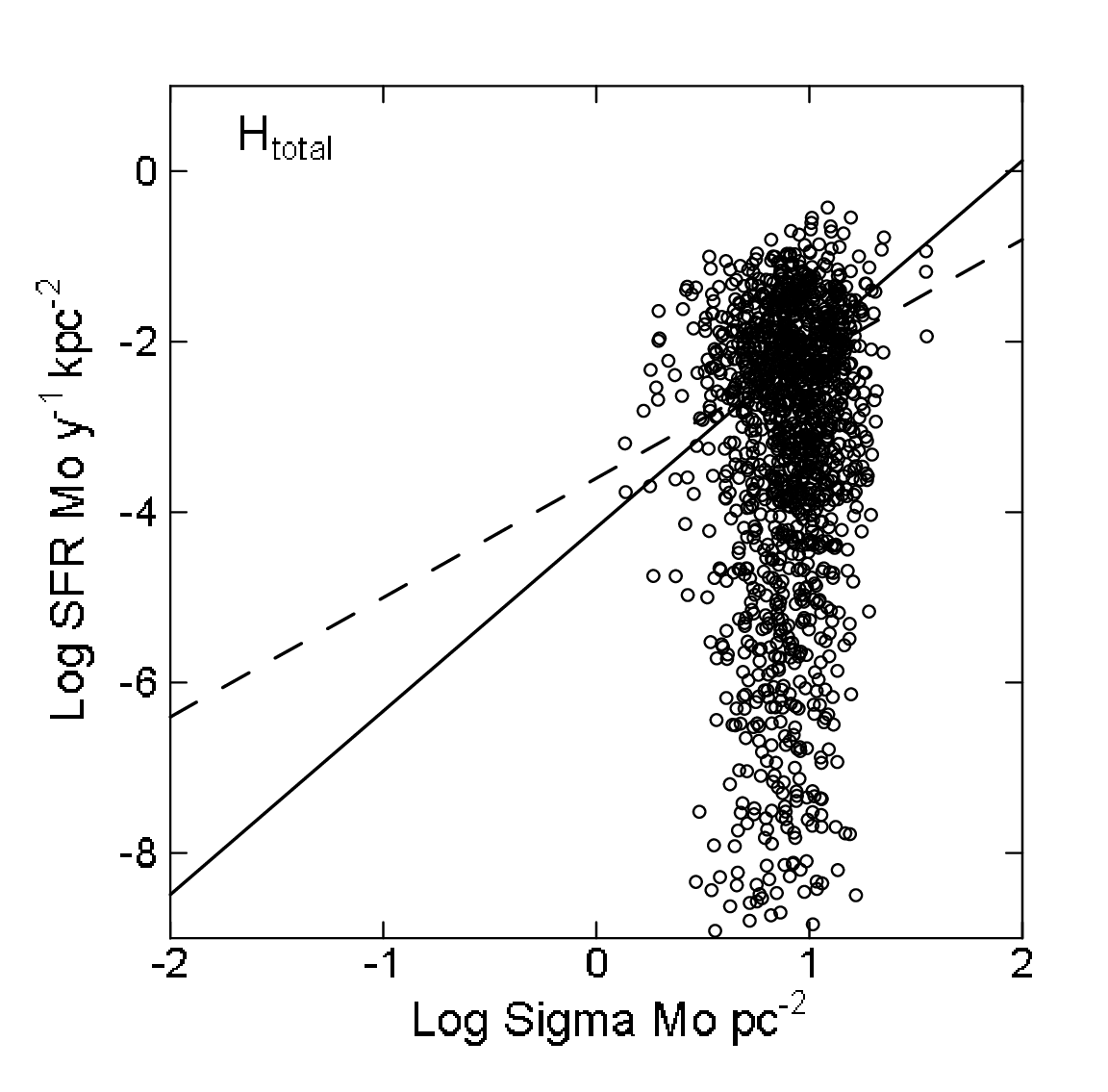}
\end{center}
\caption{[Upper panels] SF law for volume gas densities $\rhoh2$, $\rhohi$ and $\rhot$. 

[Lower panels] SF law for surface (column) gas densities $\Sigma_{\rm H_2}$, $\Sigma_{\rm HI}$ and $\Sigma_{\rm H total}$. 

Solid lines show the least-squares fitting result, and dashed line shows the Kennicutt (1998) law. }
\label{Schmidt} 
\end{figure*}

\begin{figure*} 
\begin{center}     
\includegraphics[width=7cm]{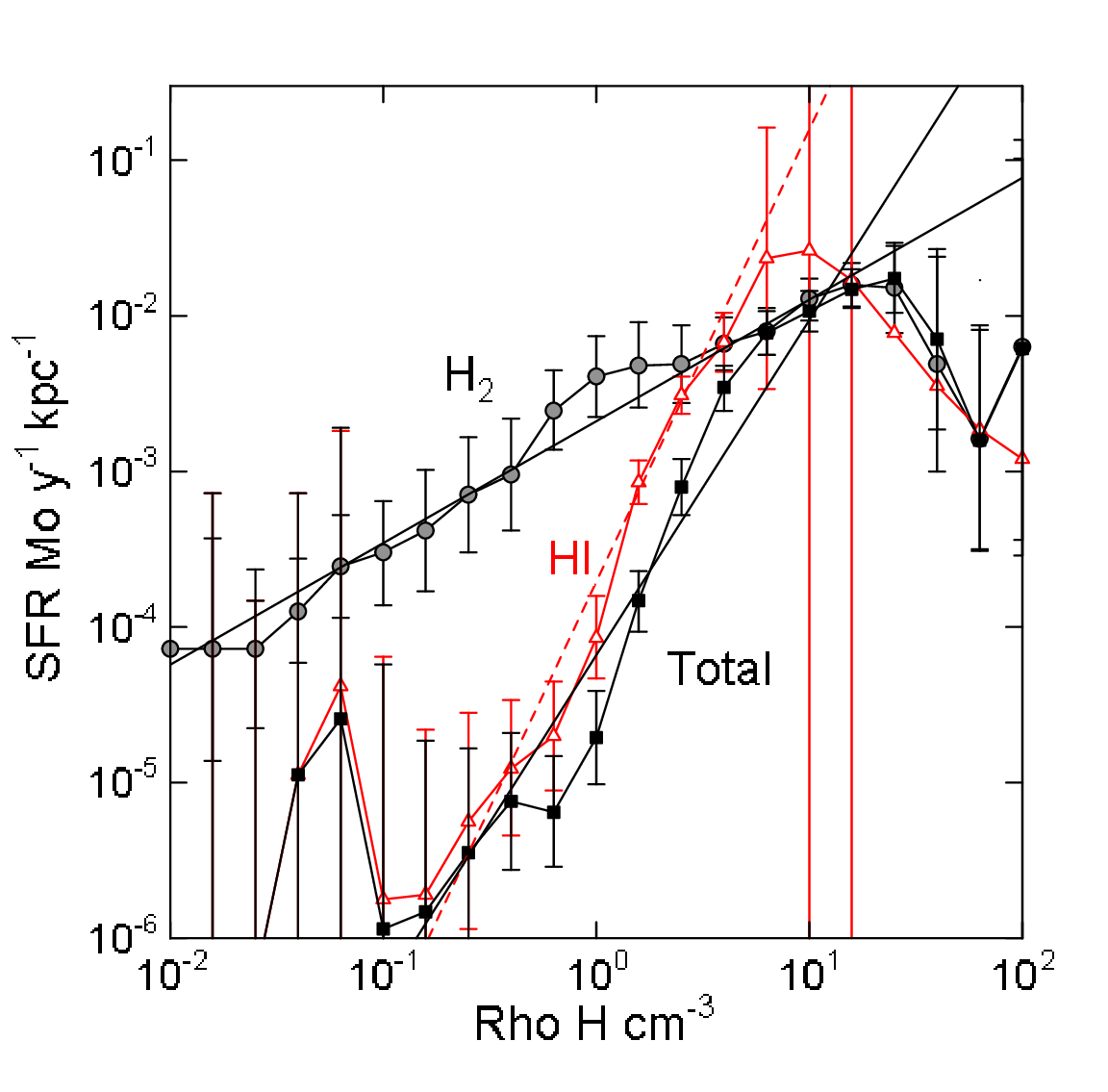}  
\includegraphics[width=7cm]{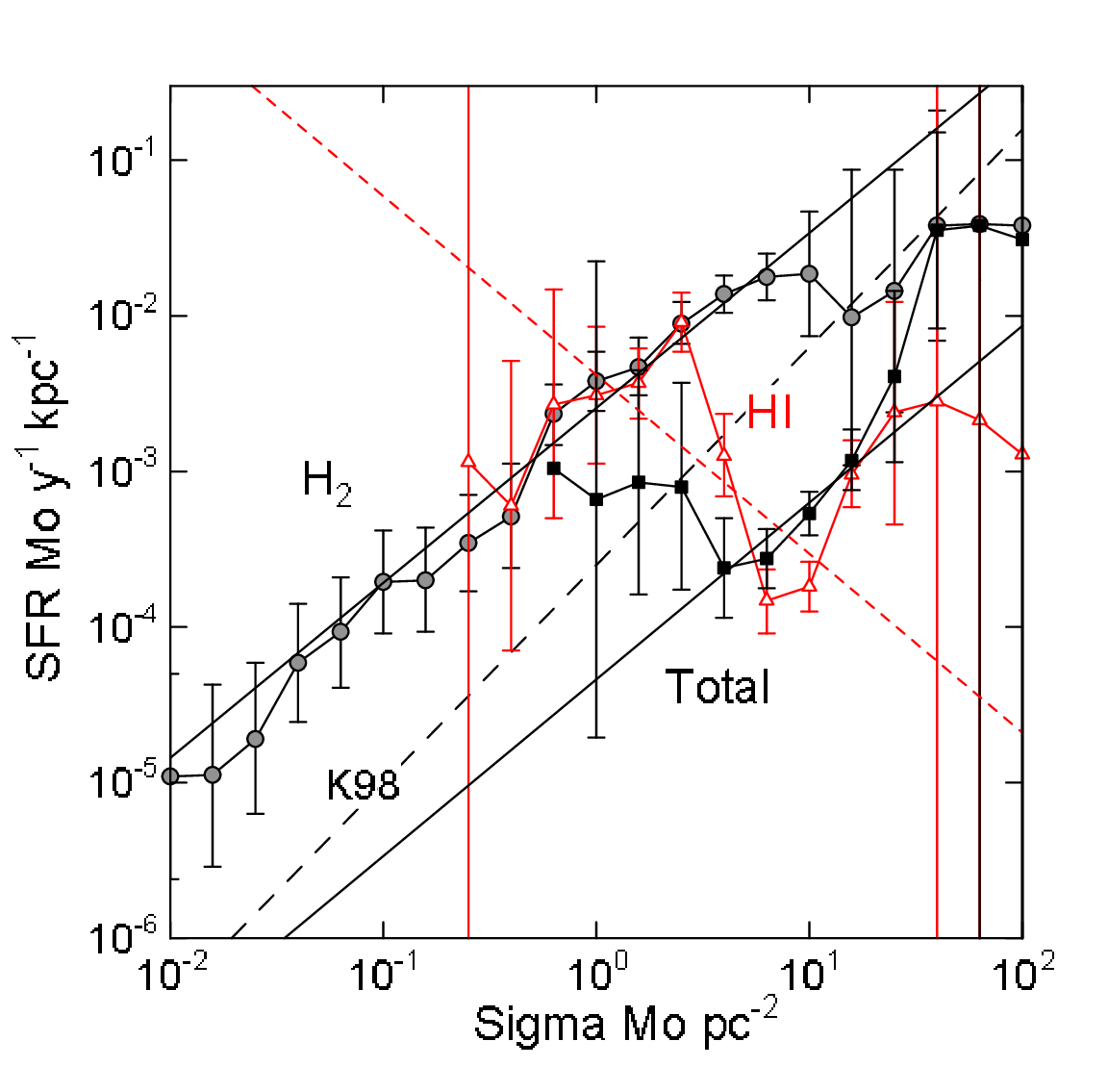}     
\end{center}
\caption{Averaged SF law and the least-squares fits, whose parameters are listed in table \ref{tabindex}. K98 stands for Kennicutt (1998) law.}
\label{CaseI}  
\end{figure*}

The distribution maps of SFR and volume and surface densities
of H2, HI and total gases in the entire Galaxy, we now analyze
the SF relation. Upper panels of figure \ref{Schmidt} shows logarithmic plots of SFR against volume densities $\rho_{\rm H_2}$, $\rho_{\rm HI}$, and $\rho_{\rm tot}$. Similar plots were obtained for the surface mass densities of gas as shown in the lower panels of figure \ref{Schmidt}. 
Figure \ref{CaseI} shows the same, but SFR for different gas phases are plotted in one panel, where values of log SFR are averaged in every 0.2 dex bins of gas density with standard errors of averaged values shown by bars. The least-squares fitting results in figure \ref{Schmidt} are inserted by straight lines.
 
The $\ssfr-\rhoh2$ and $\ssfr-\rhotot$ plots exhibits the well known Schmidt law. However, the $\ssfr-\rhohi$ plot does not show any clear SF law, but they are spread around a constant HI density of $\rhohi\sim 1-2$.  

The SF law for volume total density has a dual slopes, one for the major part at higher density of $\rhoh2>\sim 2$ \Hcc, and the other an almost vertical distribution around $\rhohi \sim 2$ \Hcc. The same behavior is observed in the $\ssfr-\Sigmahi$ and $\ssfr-\Sigmatot$ plots at $\sim 2 \Msun {\rm pc^{-2}}$. These facts indicate that there is critical density, only above which the star formation is settled.

However, no clear power-low relation was found with HI volume density. Also no clear correlation was found in the $\ssfr-\Sigmahi$ and $\ssfr-\Sigmatot$ plots. The former lack of correlation is due to the saturation of HI volume density in the galactic disk. The latter lack with HI and total surface densities is due to the thick HI disk that contributes to most of the column across the gas disk.

Although there are such structures in the plots, we here try to determine the global Schmidt law index by least-squares fitting to the data in the log-log plane.  We obtained the following SF law with the index and coefficients listed in table \ref{tabindex}.
\begin{equation} 
\log \left(\Sigma_{\rm SFR} \over \sfu \right)=\alpha\ {\rm log}\left(\rho_{\rm gas} \over {\rm H\ cm^{-3}}\right)+A,
\label{eqschrho}
\end{equation}
\begin{equation}
\log \left(\ssfr \over \sfu \right) =\beta\ {\rm log} \left(\Sigma_{\rm gas}\over {\rm \Msun~pc^{-2}}\right)+B.
\label{eqschsigma}
\end{equation}
The vertically averaged volume SFR in the SF disk of full thickness 92 pc is related to the surface SFR by
\begin{equation} 
\log \left(\vsfr \over \sfuvol \right) 
=\log \left(\ssfr \over \sfu \right)+0.967.
\label{eqschrhovol}
\end{equation} 

Meaningful fitting was obtained for $\ssfr$-$\rhoh2$, $\ssfr$-$\rhotot$, and $\ssfr$-$\Sigmah2$ as shown by the straight lines in the figure \ref{Schmidt}. 
The fitted values are listed in table \ref{tabindex} together with other automatic outputs for $\rhohi$, $\Sigmahi$, and $\Sigmatot$.

\begin{table}
\caption{SF law index and coefficient for $\ssfr$ and $\vsfr$.}
\begin{center}
\begin{tabular}{llll}  
\hline
\hline
Gas phase & $\alpha$ &$A$ for $\ssfr$ & $A$ for $\vsfr$  \\
\hline  
$\rho$ \Htwo & $   0.78 \pm    0.05$ &  $-2.68 \pm   0.05 $ &  $-1.71 \pm   0.05 $\\
$\rho$ HI    & $   2.91 \pm    0.14$ &  $ -3.71 \pm   0.05 $ &   $ -2.74 \pm   0.05 $\\
$\rho$ total & $   2.15 \pm    0.08$ &  $ -4.18 \pm   0.06 $ & $ -3.21 \pm   0.06 $\\
\hline
 &  $\beta$  &$B$   \\
\hline
$\Sigma$ \Htwo & $ 1.12 \pm   0.05 $  & $ -2.60 \pm    0.05$ \\  
$\Sigma$ HI    & $ -1.15 \pm    0.16$ & $ -2.38 \pm   0.13 $ \\
$\Sigma$ total & $  1.14 \pm    0.23$ & $ -4.34 \pm   0.22 $ \\
\hline \\
\end{tabular}
\end{center}
\label{tabindex} 
\end{table}

\section{Summary and Discussion}

\noindent{\bf Summary}

The Schmidt-Kennicutt law in the entire Milky Way was investigated using 3D distributions of HII regions, HI and \Htwo gases at spatial resolutions of $\sim 1$ kpc in the Galactic plane and a few tens of pc in the vertical direction. The vertical scale thickness of the distribution of HII regions was shown to be nearly constant with a Gaussian-fitted full thickness 92 pc, coinciding with that of the molecular gas disk. The vertically averaged volume star formation rate (SFR) $\vsfr$ is thus shown to be related to the surface SFR $\ssfr$ by
$\vsfr /[\sfuvol] =9.26\times \ssfr/[\sfu]$.
The SF law fitted by a single power law of gas density in the form of 
$\ssfr \propto \vsfr \propto \rho_{\rm gas}^\alpha$ and 
$\propto \Sigma_{\rm gas}^\beta$ 
showed indices of
$\alpha=0.78 \pm 0.05$ for $\rho_{\rm H_2}$ and 
$2.15 \pm 0.08$ for $\rho_{\rm total}$, and
$\beta=1.14\pm 0.23$ for $\Sigma_{\rm total}$, 
where $\rho$ and $\Sigma$ denote volume and surface densities, respectively. \\

\noindent{\bf Direct SF relation to \Htwo}

The tightest correlation was found
for SFR-\Htwo plot among all the other plots in figure \ref{Schmidt}. The
power-law index of $\alpha_{\rm H_2}\sim  0.78$ is about a half of the index for total gas, $\alpha_{\rm total}\sim 2.15$.
This implies that the star formation is directly related to
molecular gas through spontaneous condensation of molecular clouds. 
This is consistent with the fact that the scale thickness of the disk of HII-regions (the SF disk) is almost exactly equal to that of the molecular gas disk.\\

\noindent{\bf Indirect relation to HI}

The 3D H$_2$ and HI maps of the Galaxy (Nakanishi and Sofue 2005, 2006, 2016) exhibit significant difference in their distributions. 
H$_2$ is tightly concentrated to the galactic plane making a thin and dense molecular disk. Also, it is distributed mostly inside the molecular front at $R \sim 8$ kpc, while HI is widely distributed in the entire disk surrounding the molecular disk with thickness several times that of H$_2$ disk. On the other hand, the outer gas disk $R>\sim 8$ kpc is dominated by HI with little molecular gas, predicting little star formation.
These conditions make the correlation between SFR and surface (column) densities of HI and total gases less tight compared to that for the volume densities.\\

\noindent{\bf Steeper and/or dual power-law behavior for total gas}

The SF law for total gas is therefore diluted by inclusion of HI gas, which results in a significantly steeper index. Namely, stars are born from molecular gas, and the molecular gas is formed from HI by phase transition, intervened by various ISM processes depending on the gas density, pressure, metallicity, and radiation field, and hence on the SFR (Elmegreen 1993; Sofue and Nakanishi 2016). 
Also crucial is the almost constant HI density in the SF disk at $\rhohi\sim \rho_{\rm HI: crit}\sim 1.5$ \Hcc. We may represent the total density as 
\begin{equation}
\rhotot=\rhoh2+\rhohi\simeq \rhoh2+\rho_{\rm HI:crit}\simeq \rhoh2+{\rm const}.
\label{eqmodel}
\end{equation}  
If the SF law is expressed by a single power law, this equation leads to a steeper index than for \Htwo. 
Therefore, the observed SFR-$\rho_{\rm total}$ may be better represented by dual-slope components than by a single power: one component dot to HI gas around the critical density having no direct relation to SF, and the other due to \Htwo gas correlating directly to SF. 
On the other hand, the SFR-$\Sigma_{\rm total}$ plot is rather similar to that for HI surface density, because of the larger thickness of HI disk, whereas the power-law fit yielded $\alpha_{\rm total}\sim 1.14$, close to that of the Kennicutt law (1998), $\sim 1.4$.\\

\noindent{\bf Comparison with extragalactic values}

The Galactic SF law may be compared with that derived for external galaxies, where surface densities are used. The Galactic value, $\beta_{\rm total}\sim 1.14$, is close to the often quoted value for galaxies, $\beta_{\rm total,~galaxies}\simeq 1.4$ (Kennicutt 1998). The index $\alpha_{\rm H_2}\sim 0.78$ for molecular volume density is close to that obtained for resolved regions in extragalactic molecular disks of $\beta \sim 1.0$ (Bigiel et al. 2008). However, it is significantly smaller than that in dense gas disks of galaxies with $\sim 1-2$ (Komugi et al. 2005, 2006).\\
 
\noindent{\bf Comparison with SF law for individual SF regions}

The present spatial resolution may be too crude to discuss tighter correlations more directly related to star formation from individual molecular clouds. The observed number of HII regions is also not large enough for such  purpose.
In fact, the present result for molecular gas is much smaller than the index of $\sim 2$ obtained for individual galactic SF regions and molecular clouds by Willis et al. (2014) and similar values by Heiderman et al. (2010).

\end{document}